\documentclass[10pt,a4paper]{article} 
\usepackage[a4paper,margin=2cm]{geometry}
\usepackage{multicol} \usepackage{graphicx} \usepackage{amsmath,amssymb} \usepackage{authblk} \usepackage{abstract} \usepackage{titlesec} \usepackage{caption} 
\usepackage[colorlinks = true, linkcolor=blue, urlcolor=blue, citecolor = blue]{hyperref}
\usepackage{graphicx}
\usepackage{tabularx}
\usepackage{float}
\usepackage{subcaption}
\usepackage{color}
\usepackage{bm}
\usepackage{amssymb}
\usepackage{multirow}
\usepackage{amsmath}
\usepackage{cite}
\usepackage{hyperref}
\usepackage[normalem]{ulem}
\usepackage{amsfonts}
\usepackage{etoolbox}
\usepackage[colorlinks = true, linkcolor=blue, urlcolor=blue, citecolor = blue]{hyperref}
\usepackage{graphicx}
\usepackage{tabularx}
\usepackage{float}
\usepackage{subcaption}
\usepackage{color}
\usepackage{bm}
\usepackage{amssymb}
\usepackage{multirow}
\usepackage{amsmath}
\usepackage{cite}
\usepackage{hyperref}
\usepackage[normalem]{ulem}
\titleformat{\section} {\large\bfseries} {\thesection.}{0.5em}{} \titleformat{\subsection} {\normalsize\bfseries} {\thesubsection.}{0.5em}{} 
\title{\textbf{Investigating the mechanisms behind current redistribution and
energy transport in magnetic flux pumping scenarios: long--term saturated dynamics}} \author[1]{F. Stefanelli$^1$\thanks{Email: federico.stefanelli@ipp.mpg.de}, E. Fable$^1$, H. Zhang$^1$, H. Zohm} \author[2]{the ASDEX Upgrade team}\affil[1]{\small Max Planck Institute for Plasma Physics, 85748 Garching, Germany} \affil[2]{See author list of T. P\"{u}tterich $\textit{et al}$ 2026 $\textit{Nucl. Fusion}$, $\textbf{66}$, 116002 \hyperref[https://doi.org/10.1088/1741-4326/ae61c8]{https://doi.org/10.1088/1741-4326/ae61c8}} 
\date{November 2026} 
\begin{document} 
\maketitle \begin{abstract} {
\noindent
We develop a reduced analytical and numerical model of the magnetic flux pumping dynamics for a tokamak plasma in the presence of a core saturated quasi--interchange MHD mode. We focus on the long--time quasi--stationary regime where the safety factor $q$, the core pressure profile, and the perturbed fields interact dynamically to self--organise. The key novelty is to evaluate the perturbed radial flow consistently with the evolving equilibrium quantities. In this way we can also explain why the helical core appears to be in a stable state despite the existence of a continuous radial flow transporting flux inwards and energy outwards. The developed model is implemented in an integrated modelling framework and applied to an ASDEX Upgrade flux pumping scenario, where we can explain the observed magnetic flux dynamo. Additionally, the magnetic flux pumping dynamo is found to peak where the current drive peaks, due to the effects of local variations of the $q$ profile on the dynamo shape. This feature of the model suggests that magnetic flux pumping can be an effective mechanism to redistribute current driven by ECCD and maintain a flat $q$ profile in the core, as observed in the experiment.
} \end{abstract} 
\vspace{0.5cm} \begin{multicols}{2} 
\section{Introduction}
To predict self--consistently the evolution of core tokamak plasma kinetic (temperature, density) and magnetic (safety factor) profiles under application of heating, fuelling and current drive, one needs to model a variety of internal magnetohydrodynamic modes.\\
In a previous work \cite{Stefanelli2026} we have addressed the sawtooth cycle \cite{vonGoeler1974}, including a reduced description of the fishbone mode\cite{Mcguire1983}. 
When the core safety factor is pushed towards a flat structure and close to unity, together with a high core normalised pressure $\beta$, the quasi--interchange mode\cite{Wesson1986} can instead be destabilised. In this case, the coupling of flow perturbations and magnetic perturbations can generate a magnetic dynamo which leads to a self-organised state of the plasma known as magnetic flux pumping \cite{Petty2009,Stober2007,Krebs2017}. In this regime, the dynamo acts against current diffusion and current drive to sustain a flat and close to unity core safety factor \cite{Zhang2025}, preventing the destabilization of internal modes which would cause a sawtooth crash. \\
In \cite{Fable2026} an analysis of the quasi--interchange saturation mechanism and resulting fluctuation properties is given. However, it does not cover the long--time quasi--stationary dynamics and self--organisation of the mode. 
In general, when an internal MHD mode saturates, one would expect that the residual flow only has components tangential to the perturbed flux surfaces. This is because a radial flow (radial also with respect to the perturbed surfaces) would continue to move the magnetic field, except in the singular case in which it is perfectly balanced by non--ideal effects. However, the latter becomes difficult to invoke in cases where magnetic flux pumping, characterised by said radial flow, exists at very low values of resistivity and viscosity (as documented in, e.g. \cite{Zhang2020,Zhang2026}). \\
In this work, the persistence of a radial flow with a saturated magnetic flux surface deformation is instead explained through the effect of the nonlinearly generated higher harmonics of the mode, which balance further deformations of the magnetic flux surfaces driven by the radial flow. 
A model of the saturated flow is then derived by including self-consistently the effects of the perturbed momentum equation and the effects of the magnetic flux pumping flow itself on the equilibrium profiles.
\label{sec:intro}

\section{Physics model}
\label{sec:1}
In order to predict the amplitude of the dynamo $[\tilde{\mathbf{v}}\times \tilde{\mathbf{B}}]_\phi$ (with $[...]_\phi$ the toroidal average) driven by magnetic flux pumping, one needs a model of the perturbed radial flow amplitude $\tilde{v}_r$. To derive it, we start from a $0-\mathrm{D}$ model of the radial projection of the perturbed momentum equation for a mode with given poloidal $m$ and toroidal $n$ mode number (in this case, $m,n=1$)
\begin{equation}
    \frac{\partial \tilde{v}_r}{\partial t}=\gamma^2\xi_r-\alpha B_0^2k_\parallel^2\xi_r^3
    \label{momentum}
\end{equation}
with $\gamma$ instantaneous linear growth rate, $\xi_r$ radial deformation of the magnetic flux surfaces, $k_\parallel=\frac{1}{R_0}(\frac{1}{q}-\frac{n}{m})$, with $B_0>0\rightarrow q>0$, and $\alpha$ a parameter which depends on the shape of the profiles, which is assumed to be time-independent, and $q$ is averaged within the $q=1$ surface. 
We clarify in Section \ref{sec:2} how to model the radial profile without needing to solve a partial differential equation, as MHD would require.\\
Eq. (\ref{momentum}) is a generalisation of the nonlinear evolution equation of a quasi-interchange mode (Eq. (73) of \cite{Waelbroeck1989}). The first term on the right-hand side represents the linear drive $\gamma^2\xi_r=\frac{1}{\rho_M}(-\nabla \tilde{p}+\mathbf{\tilde{J}}\times\mathbf{B_0}+\mathbf{J_0}\times\mathbf{\tilde{B}})$, while the second term represents the nonlinear drive generated by higher harmonics $\frac{1}{\rho_M}\mathbf{\tilde{J}}\times\mathbf{\tilde{B}}=\frac{1}{\mu_0\rho_M}\mathbf{(\nabla\times\tilde{B}})\times\mathbf{\tilde{B}}$. The generation of higher harmonics has been observed in JOREK simulations \cite{Zhang2025} to lead to the saturation of a quasi-interchange mode. Due to the small inertia of the plasma, the $(\mathbf{\tilde{v}}\cdot\nabla)\mathbf{\tilde{v}}$ nonlinear term has been observed to be negligible, compared to the $\frac{1}{\rho_M}\mathbf{\tilde{J}}\times\mathbf{\tilde{B}}$ term, which is instead of the same order of magnitude of the linear drive and balances it at saturation. The viscous term $\sim\nu\nabla^2\tilde{v}$ is also neglected in the momentum equation, as it is orders of magnitudes smaller than the linear and nonlinear terms, and will be even smaller on future reactor devices that will operate at even lower dissipation levels. For this reason, viscosity does not play a significant role in the saturation of the mode in the magnetic flux pumping regime.\\ The structure of the nonlinear term in Eq. (\ref{momentum}) can be understood by employing $\tilde{B}\sim iB_0k_\parallel\xi$, valid for any harmonic, and assuming that the dominant higher harmonic is the $(2,2)$, with $\xi_{2,2}\sim\xi_{1,1}^2$, resulting in $\frac{1}{\mu_0\rho_M}\mathbf{[(\nabla\times\tilde{B}})\times\mathbf{\tilde{B}}]_{1,1}\approx\alpha \tilde{B}_{2,2}\tilde{B}_{-1,-1}\approx \alpha (B_0 k_\parallel\xi_{2,2})(B_0k_\parallel \xi_{1,1})\approx\alpha B_0^2k_\parallel^2 \xi_r^3$. $\alpha$ takes into account the relation between the perturbations in the various directions and their radial profile, as well as the factor $\frac{1}{\rho_m\mu_0}$. \\
As mentioned above, in the quasi-stationary state of magnetic flux pumping  the terms $\gamma^2\xi_r$ and $\alpha B_0^2 k_\parallel^2\xi_r^3$ are much larger than $\frac{\partial \tilde{v}_r}{\partial t}$ due to the small inertia of the plasma, but compensate each other to have at equilibrium 
\begin{equation}
    \gamma^2\xi_r-\alpha B_0^2k_\parallel^2\xi_r^3\approx0
    \label{equilibrium}
\end{equation}
One can study how deviations from the equilibrium condition affect $\tilde{v}_r$ by taking a time derivative of the momentum equation
\begin{equation}
\begin{aligned}
    \frac{\partial^2 \tilde{v}_r}{\partial t^2}=2\gamma\xi_r\frac{\partial \gamma}{\partial t}+\gamma^2\frac{\partial \xi_r}{\partial t}-2\alpha B_0^2 k_\parallel\xi_r^3 \frac{\partial k_\parallel}{\partial t}-3\alpha B_0^2 k_\parallel^2\xi_r^2\frac{\partial \xi_r}{\partial t}\\
    =2\gamma\xi_r\frac{\partial \gamma}{\partial t}-2\gamma^2\frac{\partial \xi_r}{\partial t}+2 \xi_r\frac{\gamma^2}{k_\parallel} \frac{1}{q^2R_0}\frac{\partial q}{\partial t}
\end{aligned}
    \label{momentum_derivative}
\end{equation}
where after taking the time derivative we have substituted back $\alpha=\frac{\gamma^2}{B_0^2k^2_\parallel\xi_r^2}$ since the final result will be shown to be independent of $\xi_r$.\\
To further simplify Eq. (\ref{momentum_derivative}), expressions of $\frac{\partial \gamma}{\partial t}$, $\frac{\partial \xi_r}{\partial t}$ and $\frac{\partial q}{\partial t}$ are needed. For a quasi-interchange mode \cite{Waelbroeck1989}
\begin{equation}
    \gamma\tau_A\approx a\beta_{Bussac}\approx\frac{a2\mu_0\langle r\partial_r p\rangle}{B_{p1}^2}
    \label{gamma_model}
\end{equation}
with $a$ numerical parameter, $\langle r\partial_r p\rangle=\langle p\rangle-p(r_1)$ a measure of the volume-averaged pressure gradient within the $q=1$ surface \cite{Bussac1975} and $B_{p1}\approx\frac{r_1B_\phi}{R}$ poloidal magnetic field at the $q=1$ surface. Using the linear MHD code CASTOR3D \cite{Strumberger2016}, the validity of Eq. (\ref{gamma_model}) has been also confirmed through a $\partial_r p$ scan. The field-line bending term $\sim(q-1)^2$ is neglected as it is small for $q\approx1$.
Taking a time derivative of Eq. (\ref{gamma_model}) yields
\begin{equation}
\begin{aligned}
    \frac{\partial \gamma}{\partial t}=\frac{a2\mu_0}{\tau_AB_{p1}^2}\frac{\partial\langle r\partial_r p\rangle}{\partial t}
    =\frac{\gamma}{\langle r\partial_r p\rangle}\frac{\partial\langle r\partial_r p\rangle}{\partial t}
    \end{aligned}
    \label{gamma_model_derivative}
\end{equation}
The time evolution of the pressure is given by a balance between heating $H$ and transport $\chi p$
\begin{equation}
    \frac{\partial p}{\partial t}=\frac{2}{3}(H-\frac{\partial (\chi_{0}p)}{\partial r}-\frac{\partial (\chi_{MHD}{p})}{\partial r})\approx \frac{2}{3}(\tilde{H}-\frac{\chi_{MHD}p}{r_1})
    \label{dpdt}
\end{equation}
 which includes the radial transport induced by magnetic flux pumping itself $\chi_{MHD}$ discussed in \cite{Fable2026}, both through the axisymmetric radial flow $v_{r,0}\sim\frac{\tilde{v}_\phi \tilde{B}_r}{B_0}$, which corresponds to a convection coefficient of $\frac{5}{2}v_{r,0}$, and $\tilde{v}_r\partial_r \tilde{p}$. The axisymmetric components of the energy transport arise from a $(1,1)$ perturbation coupling to a $(-1,-1)$ one, which is itself the complex conjugate of a $(1,1)$ perturbation. For this reason, the flux-surface average of the coupling of two perturbations is simply given by the multiplication of their radial profile. The total pressure entering $\gamma$ includes also the fast ion pressure $p_{f}=\frac{2}{3}\langle E_{f}\rangle n_{f}$. The time evolution of the fast-ion pressure is given by 
 \begin{equation}
     \frac{\partial p_{f}}{\partial t}= \frac{2}{3}\langle E_{f}\rangle \frac{\partial n_{f}}{\partial t}=\frac{2}{3}\langle E_{f}\rangle(S_{f}-\frac{n_{f}}{\tau_{sd}}-\frac{\chi_{MHD}n_{f}}{r_1})
     \label{pfast_evolution}
 \end{equation}
 with $S_{f}$ fast-ion density source, $\tau_{sd}$ slowing-down time and $\chi_{MHD}$ redistributes also fast-ions through $v_{r,0}$. Eq. (\ref{pfast_evolution}) can be incorporated in Eq. (\ref{dpdt}) by considering the source and sink from slowing-down in $\tilde{H}$ and the total pressure in the term $\frac{\chi_{MHD}p}{r_1}$. \\
 From JOREK \cite{Hoelzl2024} nonlinear MHD simulations it is observed that $\tilde{v}_\phi\gg \tilde{v}_r$. This property of the flow of magnetic flux pumping can be explained by evaluating the incompressibility relation along the directions parallel and perpendicular to the magnetic field lines
 \begin{equation}
     0\approx\nabla\cdot \tilde{v}_{1,1}\approx ik_\perp \tilde{v}_{r}+ik_\parallel\tilde{v}_{\parallel}
 \end{equation}
 with $k_\perp \approx \frac{1}{r_1}$ and $r_1$ minor radius of the $q=1$ surface. With $\tilde{v}_{\phi}\approx \tilde{v}_{\parallel}$ one obtains $\tilde{v}_{\phi}\approx\frac{R_0}{r_1}\frac{q}{1-q}\tilde{v}_{r}$. In JOREK simulations with fixed background transport the pressure profile is flattened by magnetic flux pumping, and due to the large value of $\tilde{v}_\phi$ the term $\frac{\tilde{v}_\phi \tilde{B}_r}{B_0}$ dominates the energy transport in the saturated phase \cite{Fable2026}. However, in reality also the background transport is affected by magnetic flux pumping, due to the modification of the temperature and density gradients, and the experimental pressure profile is not observed to flatten as much as in JOREK. For this reason, the term $\tilde{v}_r\partial_r \tilde{p}\approx-\tilde{v}_r\partial_r(\xi_r\partial_r p_0)$ can also play a significant role in the energy transport of the saturated phase.\\
The radial transport due to flux pumping can be then expressed as $\chi_{MHD}\approx\frac{5}{2}\frac{\tilde{v}_\phi \tilde{B}_r}{B_0}\approx \frac{5}{2}c\frac{\tilde{v}_r \xi_r}{r_1}$ where $c$ is a numerical parameter encapsulating the different shapes of $\tilde{v}_\phi$ and $\tilde{v}_r$, and additional terms to the energy transport such as $\tilde{v}_r\partial_r \tilde{p}$ mentioned above. The evolution of the pressure gradient reads 
\begin{equation}
    \frac{\partial \langle r\partial_r p\rangle}{\partial t}=\frac{2}{3}\langle r\partial_r(\tilde{H}-\frac{5}{2}c\frac{\tilde{v}_r \xi_r }{r_1^2}p)\rangle
\end{equation}
Similarly, for a large aspect ratio circular plasma $\frac{\partial  q}{\partial t}=-\langle\frac{q^2}{2\pi B_0r}\frac{\partial}{\partial r}(\frac{\partial\psi}{\partial t})\rangle$ with $\frac{\partial \psi}{\partial t}=2\pi\eta R_0(j_{\phi}-j_{CD}-j_{MHD})$. Here $\eta$ is the resistivity, $j_\phi$ is the total current density, $j_{CD}$ is the current density from current drive and bootstrap current, while $j_{MHD}$ represents the dynamo driven by magnetic flux pumping $\frac{1}{\eta R_0}\tilde{v}_r\tilde{B}_\theta$, expressed as additional current density. By using $\nabla\cdot \textbf{B}=0$, $\tilde{B}_\theta\approx \tilde{B}_r=iB_0k_\parallel\xi_r$ for $r\ll1$, one obtains the dynamo term $j_{MHD}=\frac{dB_0}{\eta R_0}\tilde{v}_r\xi_r k_\parallel$. $d$ is a numerical parameter encapsulating the effects of the shape of dynamo term and of additional terms contributing to the dynamo discussed in \cite{Fable2026}, such as the residual of the sum of $\tilde{v}_\theta \tilde{B}_r$ and $v_{r,0}B_{\theta,0}$, which are observed in JOREK simulations to be almost identical but of opposite sign \cite{Zhang2025, Fable2026}.\\
As far as $\frac{\partial \xi_r}{\partial t}$ is concerned, in the linear phase $\frac{\partial \xi_r}{\partial t}=\gamma\xi_r=\tilde{v}_r$, and $\frac{\partial^2 \tilde{v}_r}{\partial t^2}=-2\gamma^2\tilde{v}_r$ would result in a harmonic oscillator of frequency $\sqrt{2}\gamma$. This dynamics corresponds to the fast oscillations observed in \cite{Fable2026} in the case where the effect of pressure transport, and thus of $\frac{\partial \gamma}{\partial t}$, is neglected.
However, $\xi_r$ saturates in the nonlinear phase, which means that $\tilde{v}_r\rightarrow0$ if $\frac{\partial \xi_r}{\partial t}=\tilde{v}_r$ still holds. This is generally not the case for a nonlinearly saturated system, since higher harmonics can also generate a $(1,1)$ magnetic perturbation through nonlinear coupling and lead to the saturation of $\xi_r$ without needing $\tilde{v}_r=0$. To further elucidate this property, the relation $\frac{\partial \xi_r}{\partial t}=\tilde{v}_r$ of the linear phase can be viewed as an alternative form of the induction equation
\begin{equation}
    \frac{\partial \tilde{B}_r}{\partial t}=(\textbf{v}\times\textbf{B})_r=(\textbf{B}\cdot\nabla)\tilde{v}_r-(\textbf{v}\cdot\nabla)\tilde{B}_r
    \label{induction}
\end{equation}
with $\tilde{B}_r\sim iB_0k_\parallel \xi_r$, where the resistive current diffusion has been neglected here, since it does not play a significant role on the timescales on which the mode saturates. Since $B_{r,0}=0$, and assuming $\textbf{v}_0=0$, for linear MHD Eq. (\ref{induction}) reduces to $iB_0k_\parallel \tilde{v}_r=\frac{\partial \tilde{B}_r}{\partial t}=\gamma \tilde{B}_r=\gamma iB_0k_\parallel\xi_r \rightarrow \gamma\xi_r=\tilde{v}_r$. However, in the nonlinear case additional terms are present, such as $[(\tilde{v}_{-1,-1}\cdot\nabla)\tilde{B}_{2,2}]_{1,1}$. These terms are observed to lead to the saturation of $\tilde{B}_r$ in JOREK simulations, with the $(\tilde{\textbf{v}}\cdot\nabla)\tilde{B}$ being the dominant nonlinear term. Through this mechanism $\xi_r$ saturates while $\tilde{v}_r$ remains finite. Considering only the $(2,2)$ dominant higher harmonic with $\xi_{r2,2}\sim\xi_{r1,1}^2$, and including also resistive current diffusion for the long time-scale dynamics (even though it still provides only a small contribution due to the high $T_e$ in the core) $\frac{\partial\xi_r}{\partial t}$ can be expressed as 
\begin{equation}
    \frac{\partial\xi_r}{\partial t}=\tilde{v}_r(1-\alpha_2\xi_r^2)-\frac{\eta}{\mu_0}k^2\xi_r
    \label{induction_saturation}
\end{equation}
where $k^2=k_\parallel^2+k_\perp^2$ and $\alpha_2$ takes into account the different shapes of the terms in $(\tilde{\textbf{v}}\cdot\nabla)\tilde{B}$ and the other nonlinear terms in the induction equation. At saturation, the modes self-organise through nonlinear mode coupling into a magnetic geometry with $\tilde{v}_r(1-\alpha_2\xi_r^2)\approx0$, as observed in the JOREK simulations, with the residual balancing the small non-ideal effects, such as the resistive contribution $-\frac{\eta}{\mu_0}k^2\xi_r$, so that $\frac{\partial \xi_r}{\partial t}\approx 0$. This property is fundamental to explain the mechanism of magnetic flux pumping in the saturated phase. It is generally not valid when the profiles evolve in time, as during a transient phase where the heating power and current drive is modified, and in that case the induction equation and the momentum equation must be solved simultaneously.\\
Finally, Eq. (\ref{momentum_derivative}) reads
\begin{equation}
\begin{aligned}
    \frac{\partial^2 \tilde{v}_r}{\partial t^2}=2\gamma^2\xi_r(\frac{2}{3}\frac{\langle r\partial_r\tilde{H}\rangle}{\langle r\partial_r p\rangle} - \frac{1}{B_0k_\parallel}\langle\frac{1}{r}\frac{\partial}{\partial r}\eta (j_{\phi}-j_{CD})\rangle\\ -\xi_r(\frac{5}{3}\frac{c}{r_1^2}-\frac{d}{R_0r_1}) \tilde{v}_r)-2\gamma^2(\tilde{v}_r(1-\alpha_2\xi_r^2)-\frac{\eta}{\mu_0}k^2\xi_r)
    \label{momentum_derivative_oscillator}
\end{aligned}
\end{equation}
Eq. (\ref{momentum_derivative_oscillator}) corresponds to a forced harmonic oscillator. 
As discussed above, the term proportional to $(\tilde{v}_r(1-\alpha_2\xi_r^2)-\frac{\eta}{\mu_0}k^2\xi_r)$ can be assumed to be negligible in the saturated phase, since it is approximately $0$ for a saturated magnetic geometry. In any case, it does not affect the model on a conceptual level, since it provides an additional term linear in $\tilde{v}_r$ and one independent of $\tilde{v}_r$.\\
By setting $\frac{\partial^2 \tilde{v}_r}{\partial t^2}=0$, the value of $\tilde{v}_r$ around which the system oscillates is found.
\begin{equation}
\begin{aligned}
    \tilde{v}_r^{sat}=
    \frac{\frac{2}{3}\frac{\langle r\partial_r \tilde{H}\rangle}{\langle r\partial_r p\rangle}-\frac{1}{B_0k_\parallel} \langle\frac{1}{ r}\frac{\partial}{\partial r}\eta (j_{\phi}-j_{CD})\rangle}{\xi_r(\frac{5}{3}\frac{c}{r_1^2}-\frac{d}{R_0r_1})}
\end{aligned}
\label{vrsat}
\end{equation}
The system is stable and oscillates around $\tilde{v}_r^{sat}$ for $\tilde{v}_r^{sat}>0$, while it is unstable for $\tilde{v}_r^{sat}<0$. The necessary conditions to have magnetic flux pumping are then that $v_r^{sat}>0$ and that the mode is linearly unstable, $\gamma>0$. On the long-time scale, the oscillations are damped by viscosity, while the equilibrium $v_r^{sat}$ around which the system oscillates is not affected by viscosity, as long as its contribution to the momentum balance is low and the equilibrium condition $\gamma^2\xi_r=\alpha B_0^2 k_\parallel \xi_r^3$ holds. If viscosity is high and its contribution to the momentum balance is significant, $\alpha$ depends also on $v_r$ and nonlinear terms in $v_r$ would appear in Eq. (\ref{momentum_derivative_oscillator}). It is also worth noting that, since all MHD-driven fluxes scale as $\tilde{v}_r\xi_r$, they are independent of the amplitude of $\xi_r$, due to the scaling of $\tilde{v}_r$ as $1/\xi_r$. This property is no longer valid if the term $(\tilde{v}_r(1-\alpha_2\xi_r^2)-\frac{\eta}{\mu_0}k^2\xi_r)$, arising from $\frac{\partial \xi_r}{\partial t}$ is not negligible in Eq. (\ref{momentum_derivative_oscillator}), as, for instance, during transients. In that case, a model of the amplitude of $\xi_r$ is also required, and this extension of the model is left for future work.\\
$\tilde{v}_r^{sat}$ represents the perturbed flow velocity which acts as restoring force, preventing the system to deviate from its saturated state through variations of the pressure and $q$ profiles due to heating and current drive. Without this flow, current diffusion would bring $q$ far from unity and turn the mode from a quasi-interchange to a kink, leading to a sawtooth crash. On the other hand, if $q$ is maintained flat and close to unity through external current drive, $\tilde{v}_r^{sat}$ can be $0$, in agreement with the nonlinear MHD simulations results of \cite{Adulsiriswad2026}, with a magnetic structure which is, in principle, indistinguishable from the one of the magnetic flux pumping case.\\
Since the denominator is larger than $0$ for $c$ and $d$ of the same order of magnitude, the system is stable if the heating power $\tilde{H}$ is strong enough to overcome the current peaking due to current diffusion and current drive. This property of the model is consistent with the experimental results of \cite{Burckhart2023}, where it was shown how increasing NBI power in ASDEX Upgrade would allow entering the magnetic flux pumping regime, while increasing on-axis ECCD would cause the breaking of magnetic flux pumping and a transition to the sawtooth regime. It is also consistent with the simulation results of heat conductivity scans performed with JOREK, where a lower background conductivity (and thus a higher $\tilde{H}=H-\frac{\partial\chi_0p}{\partial r}$) would result in a higher $\tilde{v}_r^{sat}$. Finally, it is consistent with the results of resistivity scans in JOREK simulations \cite{Zhang2026}, where a lower resistivity, and thus a lower $\frac{\partial q}{\partial t}$ due to the higher current diffusion time, would cause a transition from the sawtooth regime to flux pumping.
\section{Simulation results}
\label{sec:2}
\begin{figure}[H]
    \centering
        \centering
        \includegraphics[clip,trim=0.1cm 0.1cm 0.1cm 1.2cm, width=0.49\textwidth]{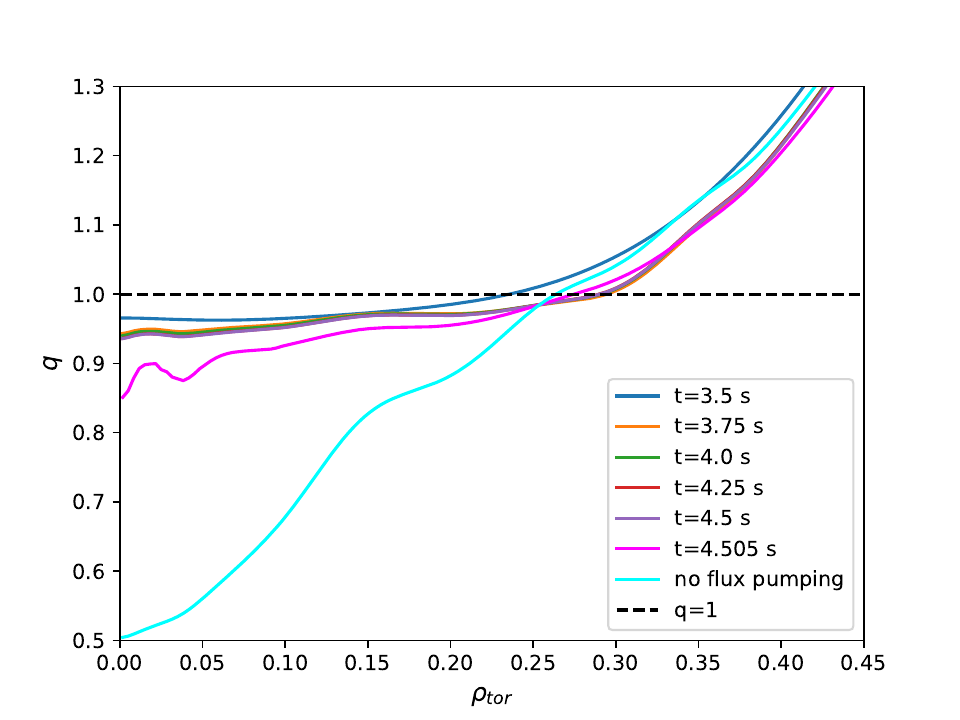}
        \caption{Simulated $q$ profile at different time steps. The flux pumping phase goes from $3.5\ \mathrm{s}$ to $4.5\ \mathrm{s}$, while it is lost at $4.5\ \mathrm{s}$, when an additional gyrotron is turned on. The cyan curve shows the $q$ profile after $1\ \mathrm{s}$ of simulation without flux pumping.}  
        \label{q_profile}
\end{figure}

The results of Section \ref{sec:1} are tested in a current diffusion simulation of the AUG magnetic flux pumping discharge, using the ASTRA integrated modelling framework \cite{Tardini2026}. The simulated discharge is $\#36663$, which has been extensively studied in \cite{Burckhart2023}. Magnetic flux pumping is reached in the experiment between $t=3.5\ \mathrm{s}$ and $t=4.5\ \mathrm{s}$, which is also the time window of the simulation. Following the approach of \cite{Fable2026}, the shape of the dynamo $[\textbf{v}\times\textbf{B}]_\phi$ is given by
\begin{equation}
\begin{aligned}
    [\textbf{v}\times\textbf{B}]_\phi\approx \tilde{v}_r\tilde{B}_\theta\approx \tilde{v}_r(-B_0k_\parallel\frac{\partial (r\xi_r)}{\partial r}+(\frac{B_{\theta,0}}{r}-\frac{\partial B_\theta}{\partial r})\xi_r)
\end{aligned}
\label{dynamo_astra}
\end{equation}
We observed that the shapes of the perturbations predicted by linear MHD qualitatively agree with the ones in the saturated phase predicted by JOREK. For this reason, the shapes of $\tilde{v}_r$ and $\xi_r$ are obtained from CASTOR3D \cite{Strumberger2016}, and Eq. (\ref{vrsat}) is used to set the amplitude of $\tilde{v}_r$. As mentioned in Section \ref{sec:1}, since $\tilde{v}_r^{sat}$ scales as $1/\xi_r$, $\tilde{v}_r\tilde{B}_\theta$ is independent of the amplitude of $\xi_r$. \\
The dynamo is included in the simulation as additional current drive $\frac{\tilde{v}_r\tilde{B}_\theta}{\eta R_0}$, and the pressure profile is kept fixed to the experimental one, which is stationary during the magnetic flux pumping phase of the discharge. The parameter $c$ is estimated by taking the radial profiles of $v_{r,0}$ and $\tilde{v}_r\partial_r\tilde{p}$, calculating their contribution to $\frac{\partial \gamma}{\partial t}$, and dividing it by $\frac{1}{r^2}$, which results into $\frac{5}{3}c\approx11$. Similarly, $d$ is estimated by calculating the contribution of the dynamo term of Eq. (\ref{dynamo_astra}) to $\frac{\partial q}{\partial t}$ and dividing it by $\frac{1}{R_0r}$, resulting into $d\approx50$. The simulation starts with a $q$ profile with low shear in the core and $q_0\approx0.97$, which is compatible with the destabilisation of quasi-interchange modes. \\
The background transport which enters $\tilde{H}$ is calculated with a power balance, resulting in $\frac{2}{3}\frac{\langle r\partial_r\tilde{H}\rangle}{\langle r\partial_r p\rangle}\approx20\ \mathrm{s^{-1}}$ and $\frac{1}{B_0k_\parallel} \langle\frac{1}{ r}\frac{\partial}{\partial r}\eta (j_{\phi}-j_{CD})\rangle\approx 7\ \mathrm{s^{-1}}$.
\begin{figure}[H]
    \centering
        \centering
        \includegraphics[clip,trim=0.05cm 0.05cm 0.05cm 0.05cm, width=0.49\textwidth]{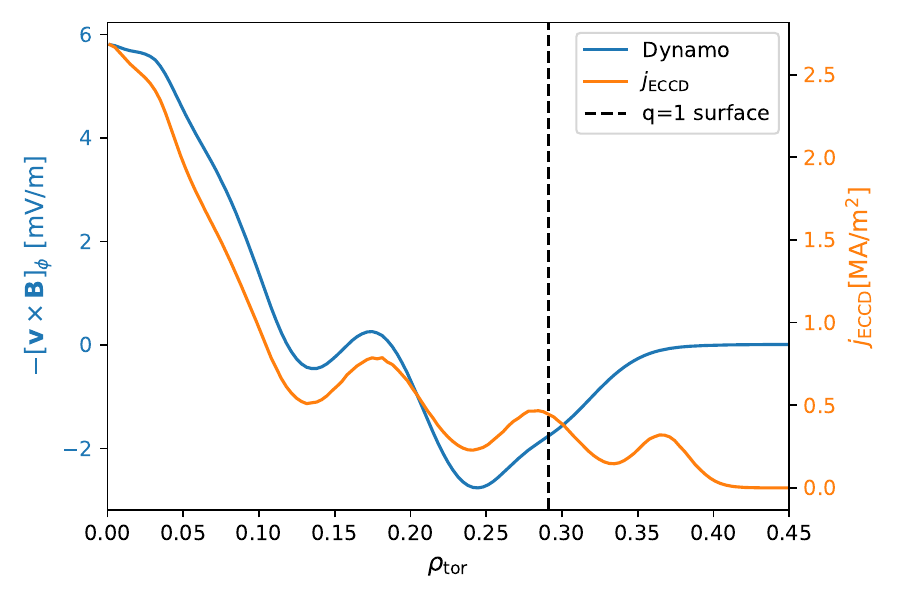}
        \caption{Dynamo (blue) and Electron Cyclotron Current Drive (orange) at $t=4.0\ \mathrm{s}$.}  
        \label{dynamo_eccd}
\end{figure}
Figure \ref{dynamo_eccd} shows the dynamo profile, as well as the profile of the Electron Cyclotron current drive (ECCD). Due to the local variation of $q$ caused by ECCD, the dynamo peaks where the current drive peaks. This property has been confirmed also with dedicated JOREK simulations. The term in the dynamo $\sim(\frac{B_{\theta,0}}{r}-\frac{\partial B_\theta}{\partial r})\xi_r\tilde{v}_r\approx\frac{s}{q}\frac{B_0}{R_0}\xi_r\tilde{v}_r$, with $s=\frac{r}{q}\frac{d q}{d r}$ magnetic shear, tends instead to flatten the $q$ profile within the $q=1$ surface. These two features suggest that magnetic flux pumping can be effective in redistributing ECCD and thus allow to drive ECCD in the plasma center, where it is the most efficient\cite{Prater2004}, while retaining a flat $q$ profile through the effects of the dynamo. The predicted dynamo has an amplitude of $\approx6\ \mathrm{mV/m}$, consistent with the one obtained with JOREK \cite{Zhang2025} and with the loop voltage deficit calculated in \cite{Burckhart2023} to recover the experimental $q$ profile.
The $k_\parallel$ dependence of Eq. (\ref{dynamo_astra}) provides a negative feedback for the saturated amplitude of the dynamo, which reaches $\approx6\ \mathrm{mV/m}$ also if the parameters $c$ and $d$ are varied.
This property makes the model robust against the choice of parameters.\\
Figure \ref{q_profile} shows the $q$ profile as function of time, which, in the flux pumping phase, quickly saturates to an average value of $\approx0.97$ within the $q=1$ surface, consistent with the JOREK results. The shear at the $q=1$ surface is less than $0.2$, which is expected to be too low to cause a sawtooth crash through a resistive kink\cite{Stefanelli2026}. Without magnetic flux pumping, the value of $q$ on axis is found to reach $\approx0.5$ after $1\ \mathrm{s}$ of simulation, as show in Figure \ref{q_profile}.\\
Regarding the amplitude of the radial flow $\tilde{v}_r^{sat}$, it is of $\approx4.5\ \mathrm{m/s}$ between $t=3.5\ \mathrm{s}$ and $t=4.5\ \mathrm{s}$, consistent with the values found in JOREK simulations of discharge $\#36663$. At $t=4.5\ \mathrm{s}$ another gyrotron is added to the ECCD in the experiment, and magnetic flux pumping is lost. In the simulation, when the additional gyrotron is included $\tilde{v}_r^{sat}\approx -5\ \mathrm{m/s}$, where it has been assumed that the system is in a new saturated phase after the transient phase due to the variation of heating and current drive. Being $\tilde{v}_r^{sat}$ negative, magnetic flux pumping is unstable and drives $q$ significantly below $1$. In just $5\  \mathrm{ms}$ the value of $q$ on-axis reaches $0.85$, as shown in Figure \ref{q_profile}, which is compatible with the transition to a kink and the loss of magnetic flux pumping.

\section{Conclusions}
A reduced model of the perturbed flow in the saturated phase of magnetic flux pumping is derived, which explains the origin of this persistent flow even when the magnetic geometry saturates. The flow arises as restoring force against the external heating and current drive, which would lead the system away from steady state. Depending on the sign of this flow, the model allows also to determine whether the magnetic flux pumping state is stable.\\
The model is tested in an integrated modelling framework and validated on a AUG magnetic flux pumping discharge. The resulting dynamo is consistent with the loop voltage deficit observed in the experiment and it allows the simulated safety factor to saturate to a flat profile close to unity. When an additional gyrotron is added, the model predicts an unstable flux pumping state, consistently with the loss of flux pumping in the experiment. Furthermore, due to the local variation of the $q$ profile, the shape of the dynamo peaks where current drive peaks, suggesting that magnetic flux pumping can be particularly effective in redistributing the current from ECCD. Finally, the model predicts that a lower current diffusion time tends to destabilise flux pumping. This result is consistent with the findings of resistivity scans performed with nonlinear MHD simulations \cite{Zhang2026}. Conversely, the longer current diffusion times expected in future fusion reactors, owing to their higher core $T_e$, are expected to favour the sustainment of magnetic flux pumping.\\
Simulations including the self-consistent evolution of the pressure profile are currently ongoing and their results will be presented in a future publication.
Finally, future work includes the validation on more discharges and machines and predictions for future machines.
\label{sec:conclusions}

\section*{Data availability statement}

The data supporting the results of this work is available from the authors upon reasonable request.

\section*{Acknowledgements}
We thank Prof S. Günter (MPG-IPP) for useful discussions on the
saturation mechanism of internal MHD modes.\\ This work has been carried out within the framework of the EUROfusion Consortium, funded by the European Union via the Euratom Research and Training Programme (Grant Agreement No 101052200-EUROfusion). Views and opinions expressed are however those of the authors only and do not necessarily reflect those of the European Union or the European Commission. Neither the European Union nor the European Commission can be held responsible for them. \\
The simulation work was partially carried out using supercomputer resources, in the Pitagora HPC at Bologna, Italy and Plasma Simulator at Rokkasho, Japan provided under the EU-JA Broader Approach collaboration and the Computational Simulation Centre of International Fusion Energy Research Centre (IFERC-CSC). The simulation work also utilized the TOK cluster at the Max Planck Institute for Plasma Physics.

\AtBeginEnvironment{thebibliography}{\small}
\bibliographystyle{unsrt}
\bibliography{refs.bib}

@article{Stefanelli2026,
  title = {Integrated modelling of sawtooth cycles and fishbone oscillations in ASDEX Upgrade plasmas},
  volume = {66},
  ISSN = {1741-4326},
  url = {http://dx.doi.org/10.1088/1741-4326/ae6e55},
  DOI = {10.1088/1741-4326/ae6e55},
  number = {7},
  journal = {Nuclear Fusion},
  publisher = {IOP Publishing},
  author = {Stefanelli,  F. and Fable,  E. and Angioni,  C. and Bergmann,  M. and Brioschi,  D. and Gude,  A. and Igochine,  V. and Lauber,  P. and Samoylov,  O. and Stucchi,  R. and Tardini,  G. and Weiland,  M. and Zohm,  H. and },
  year = {2026},
  pages = {076003}
}

@unpublished{Fable2026,
  title = {\textit{Investigating the mechanisms behind magnetic and kinetic transport in flux pumping scenarios: linear and early saturation phases}},
  volume = {},
  ISSN = {},
  url = {},
  DOI = {},
  number = {},
  journal = {Nuclear Fusion},
  publisher = {IOP Publishing},
  author = {Fable, E. and Zhang, H. and Stefanelli, F. and Zohm, H },
  year = {2026},
  note = {Submitted to Nuclear Fusion},
  month = nov 
}

@article{Waelbroeck1989,
  title={Nonlinear growth of the quasi-interchange instability},
  author={Waelbroeck, FL and others},
  journal={Physics of Fluids B Plasma Physics},
  volume={1},
  number={3},
  pages={499},
  year={1989}
}

@article{tardini2026,
  title={ASTRA-8: a modern framework for transport analysis and modelling in fusion devices},
  author={Tardini, G and Fable, E and Angioni, C and Bergmann, M and Fajardo, D and Luda, T},
  journal={Plasma Physics and Controlled Fusion},
  volume={68},
  number={6},
  pages={065024},
  year={2026},
  publisher={IOP Publishing}
}

@article{Strumberger2016,
  title={CASTOR3D: linear stability studies for 2D and 3D tokamak equilibria},
  author={Strumberger, E and G{\"u}nter, S},
  journal={Nuclear Fusion},
  volume={57},
  number={1},
  pages={016032},
  year={2016},
  publisher={IOP Publishing}
}

@article{Burckhart2023,
  title={Experimental evidence of magnetic flux pumping in ASDEX upgrade},
  author={Burckhart, A and Bock, A and Fischer, R and P{\"u}tterich, T and Stober, J and G{\"u}nter, S and Gude, A and Hobirk, J and H{\"o}lzl, M and Igochine, V and others},
  journal={Nuclear Fusion},
  volume={63},
  number={12},
  pages={126056},
  year={2023},
  publisher={IOP Publishing}
}

@article{Petty2009,
  title={Magnetic-flux pumping in high-performance, stationary plasmas with tearing modes},
  author={Petty, CC and Austin, ME and Holcomb, CT and Jayakumar, RJ and La Haye, RJ and Luce, TC and Makowski, MA and Politzer, PA and Wade, MR},
  journal={Physical review letters},
  volume={102},
  number={4},
  pages={045005},
  year={2009},
  publisher={APS}
}

@article{Stober2007,
  title={The role of the current profile in the improved H-mode scenario in ASDEX Upgrade},
  author={Stober, J and Sips, ACC and Angioni, C and Forest, CB and Gruber, O and Hobirk, J and Horton, LD and Maggi, CF and Maraschek, M and Martin, P and others},
  journal={Nuclear Fusion},
  volume={47},
  number={8},
  pages={728--737},
  year={2007}
}

@article{vonGoeler1974,
  title={Studies of internal disruptions and m= 1 oscillations in tokamak discharges with soft—x-ray tecniques},
  author={Von Goeler, Stodiek and Stodiek, W and Sauthoff, N},
  journal={Physical Review Letters},
  volume={33},
  number={20},
  pages={1201},
  year={1974},
  publisher={APS}
}

@article{Mcguire1983,
  title={Study of high-beta magnetohydrodynamic modes and fast-ion losses in PDX},
  author={McGuire, K and Goldston, R and Bell, M and Bitter, M and Bol, K and Brau, K and Buchenauer, D and Crowley, T and Davis, S and Dylla, F and others},
  journal={Physical Review Letters},
  volume={50},
  number={12},
  pages={891},
  year={1983},
  publisher={APS}
}

@article{Adulsiriswad2026,
  title={Helical core formation and MHD stability in ITER-scale plasmas with fusion-born alpha particles},
  author={Adulsiriswad, P and Bierwage, A and Yagi, M},
  journal={Nuclear Fusion},
  volume={66},
  number={4},
  pages={046032},
  year={2026},
  publisher={IOP Publishing}
}

@article{Prater2004,
  title={Heating and current drive by electron cyclotron waves},
  author={Prater, Ronald},
  journal={Physics of plasmas},
  volume={11},
  number={5},
  pages={2349--2376},
  year={2004},
  publisher={American Institute of Physics}
}

@article{Wesson1986,
  title={Sawtooth oscillations},
  author={Wesson, John A},
  journal={Plasma physics and controlled fusion},
  volume={28},
  number={1A},
  pages={243--248},
  year={1986}
}

@article{Zhang2020,
  title={Sawtooth relaxation oscillations, nonlinear helical flows and steady-state m/n= 1 magnetic islands in low-viscosity tokamak plasma simulations},
  author={Zhang, W and Ma, ZW and Porcelli, F and Zhang, HW and Wang, X},
  journal={Nuclear Fusion},
  volume={60},
  number={9},
  pages={096013},
  year={2020},
  publisher={IOP Publishing}
}

@article{Krebs2017,
  title={Magnetic flux pumping in 3D nonlinear magnetohydrodynamic simulations},
  author={Krebs, I and Jardin, SC and G{\"u}nter, S and Lackner, K and Hoelzl, M and Strumberger, E and Ferraro, N},
  journal={Physics of Plasmas},
  volume={24},
  number={10},
  year={2017},
  publisher={AIP Publishing}
}

@article{Zhang2025,
  title={Quantitative magnetohydrodynamic modelling of flux pumping in ASDEX Upgrade},
  author={Zhang, Haowei and H{\"o}lzl, Matthias and Krebs, Isabel and Burckhart, Andreas and Bock, Alexander and G{\"u}nter, Sibylle and Igochine, Valentin and Lackner, Karl and Ramasamy, Rohan and Zohm, Hartmut and others},
  journal={Nuclear Fusion},
  volume={65},
  number={6},
  pages={066001},
  year={2025},
  publisher={IOP Publishing}
}

@article{Hoelzl2024,
  title={Non-linear MHD modelling of transients in tokamaks: a review of recent advances with the JOREK code},
  author={Hoelzl, Matthias and Huijsmans, Guido TA and Artola, Francisco Javier and Nardon, Eric and Becoulet, Marina and Schwarz, N and Cathey, Andr{\'e}s and Pamela, SJP and Aleynikova, K and Antlitz, F and others},
  journal={Nuclear Fusion},
  volume={64},
  number={11},
  pages={112016},
  year={2024},
  publisher={IOP Publishing}
}

@article{Bussac1975,
  title={Internal kink modes in toroidal plasmas with circular cross sections},
  author={Bussac, MN and Pellat, R and Edery, D and Soule, JL},
  journal={Physical Review Letters},
  volume={35},
  number={24},
  pages={1638},
  year={1975},
  publisher={APS}
}

@article{Zhang2026,
  title={Flux pumping and bifurcated relaxations of helical core in 3D magnetohydrodynamic modelling of ASDEX Upgrade plasmas},
  author={Zhang, H and Hoelzl, M and Krebs, I and Burckhart, A and Bock, A and Guenter, S and Igochine, V and Lackner, K and Bonfiglio, D and Fable, E and others},
  journal={Nuclear Fusion},
  volume={66},
  number={7},
  pages={076027},
  year={2026},
  publisher={IOP Publishing}
}
 \end{multicols} 
\end{document}